\documentclass[letter]{aa} 

\usepackage{graphicx}
\usepackage{txfonts}
\begin{document}

\title{The variation of tidal dissipation in the convective envelope\\ of low-mass stars along their evolution}

\author{
S. Mathis\inst{1}
}

\institute{Laboratoire AIM Paris-Saclay, CEA/DSM - CNRS - Universit\'e Paris Diderot, IRFU/SAp Centre de Saclay, F-91191 Gif-sur-Yvette Cedex, France
\email{stephane.mathis@cea.fr} 
}

\date{Received ... / accepted ...}

\abstract 
{Since 1995, more than 1500 exoplanets have been discovered around a large diversity of host stars (from M- to A-type stars). Tidal dissipation in stellar convective envelopes is a key actor that shapes the orbital architecture of short-period systems.}
{Our objective is to understand and evaluate how tidal dissipation in the convective envelope of low-mass stars (from M to F types) depends on their mass, evolutionary stage and rotation.}
{Using a simplified two-layer assumption, we compute analytically the frequency-averaged tidal dissipation in their convective envelope. This dissipation is due to the conversion into heat of the kinetic energy of tidal non wave-like/equilibrium flow and inertial waves because of the viscous friction applied by turbulent convection. Using grids of stellar models allows us to study the variation of the dissipation as a function of stellar mass and age on the Pre-Main-Sequence and on the Main-Sequence for stars with masses spanning from $0.4$ to $1.4M_{\odot}$.}
{During their Pre-Main-Sequence, all low-mass stars have an increase of the frequency-averaged tidal dissipation for a fixed angular velocity in their convective envelope until they reach a critical aspect and mass ratios (respectively $\alpha=R_{\rm c}/R_{\rm s}$ and $\beta=M_{\rm c}/M_{\rm s}$ where $R_{\rm s},M_{\rm s},R_{\rm c}$ and $M_{\rm c}$ are the star's radius and mass and its radiative core's top radius and mass). Next, the dissipation evolves on the Main Sequence to an asymptotic value that {becomes} maximum for {\bf $0.6M_{\odot}$} K-type stars and that decreases by several orders of magnitude with increasing stellar mass. Finally, the rotational evolution of low-mass stars strengthens the importance of tidal dissipation during the Pre-Main-Sequence for star-planet and multiple star systems.}
{As shown by observations, tidal dissipation in stars varies over several orders of magnitude as a function of stellar mass, age and rotation. We demonstrate that: i) it reaches a maximum value on the Pre-Main-Sequence for all stellar masses; ii) on the Main-Sequence and at fixed angular velocity, it {becomes} maximum for {\bf $0.6M_{\odot}$} K-type stars and it decreases with {increasing} mass.}


\titlerunning{The variation of tidal dissipation in the convective envelope of low-mass stars along their evolution}
\authorrunning{S. Mathis}

\maketitle


\section{Introduction and context}

Since 1995, more than 1500 exoplanets have been discovered around a large diversity of host stars \citep[e.g.][]{Perryman2011}. As of today, their mass range spreads from M red dwarfs to intermediate-mass A-type stars. In this context, tidal dissipation in the host star has a strong impact on the orbital configuration of short-period systems. Indeed, the dissipation in rotating turbulent convective envelopes of low-mass stars is believed to play a key-role for tidal migration, orbits' circularization and spins' alignment \citep[e.g.][and references therein for hot-Jupiter systems]{Albrechtetal2012,Lai2012,Ogilvie2014}. In these regions, this is the turbulent friction acting on tidal flows, which dissipates their kinetic energy into heat, that drives the tidal evolution of star-planet systems. In stellar convection zones, tidal flows are constituted by a large-scale non-wave like/equilibrium flow driven by the hydrostatic adjustment of the stellar structure because of the presence of the planetary/stellar companion \citep{Zahn1966b,RMZ2012} and the dynamical tide constituted by inertial waves, their restoring force being the Coriolis acceleration \cite[e.g.][]{OL2007}. In this framework, both the structure and rotation of stars strongly varies along their evolution \cite[e.g.][]{Siessetal2000,GB2013,GB2015}. Moreover, as reported by \cite{Ogilvie2014}, observations of star-planet and binary star systems show that tidal dissipation varies over several orders of magnitude. Therefore, one of the key questions that must be addressed is: {\it how does tidal dissipation in the convective envelope of low-mass stars varies as a function of stellar mass, evolutionary stage and rotation?}

In this work, we study the variations of the frequency-averaged tidal dissipation in stellar convective envelopes as a function of the mass, the age and the rotation of stars. In sec. 2, we introduce the assumptions and the formalism that allows us to analytically evaluate this quantity as a function of the structure and rotation of stars \citep{Ogilvie2013}. In sec. 3, we compute it as a function of stellar mass and evolutionary stage at fixed angular velocity using grids of stellar models for stars from $0.4$ to $1.4M_{\odot}$. Next, we discuss the impact of the rotational evolution of stars. In sec. 4, we present our conclusions and the perspectives of this work.

\section{Tidal dissipation modelling}

\begin{figure}[!t]
\centering
\includegraphics[width=0.25\textwidth]{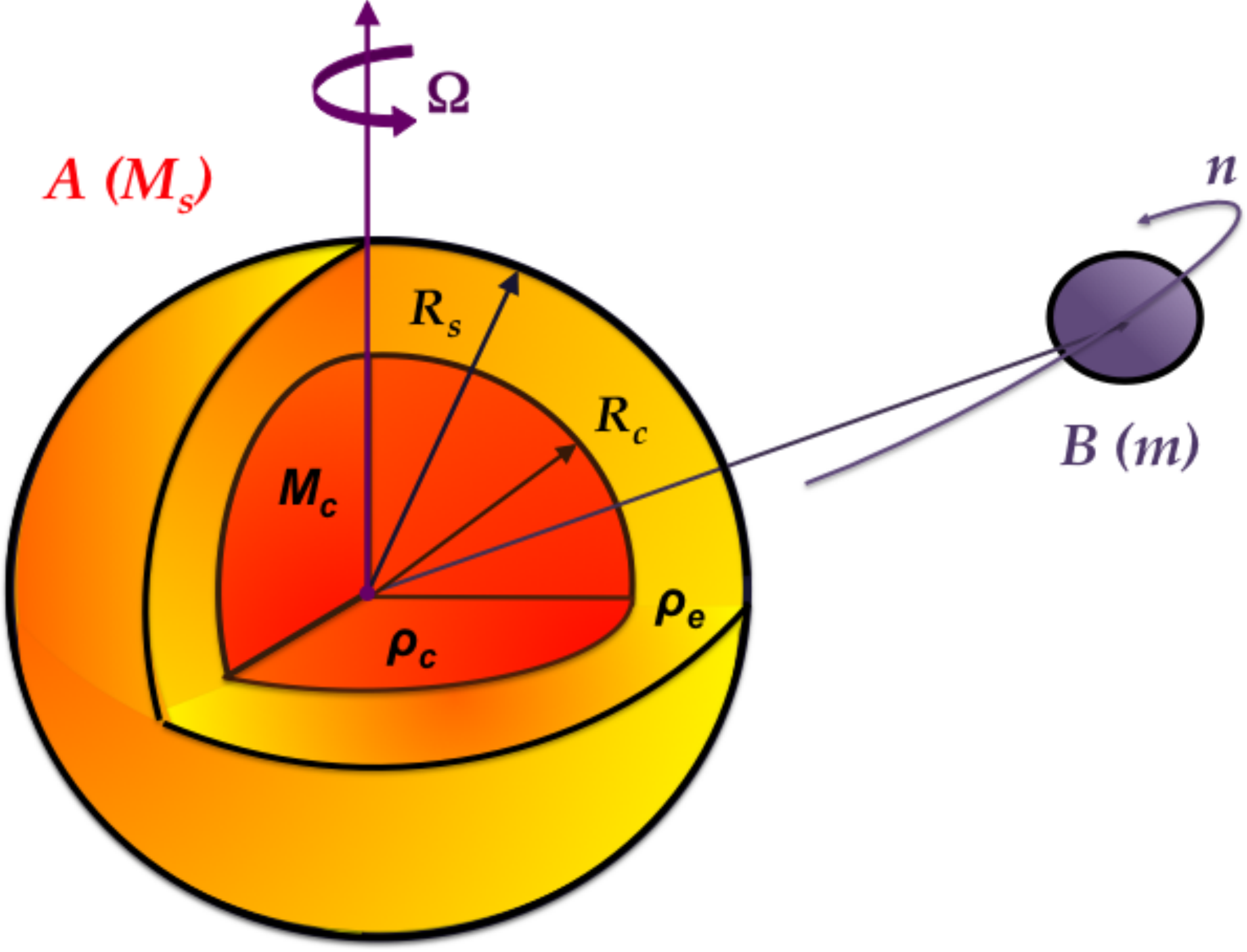}
\caption{Two-layer low-mass star A of mass $M_{\rm s}$ and mean radius $R_{\rm s}$ and point-mass tidal perturber B of mass $m$ orbiting with a mean motion $n$. The radiative core of radius $R_{\rm c}$, mass $M_{\rm c}$ and density $\rho_{\rm c}$ is surrounded by the convective envelope of density $\rho_{\rm e}$.}
\label{SetUp}
\end{figure}

\begin{figure*}[!t]
\begin{center}
\includegraphics[width=0.35\linewidth]{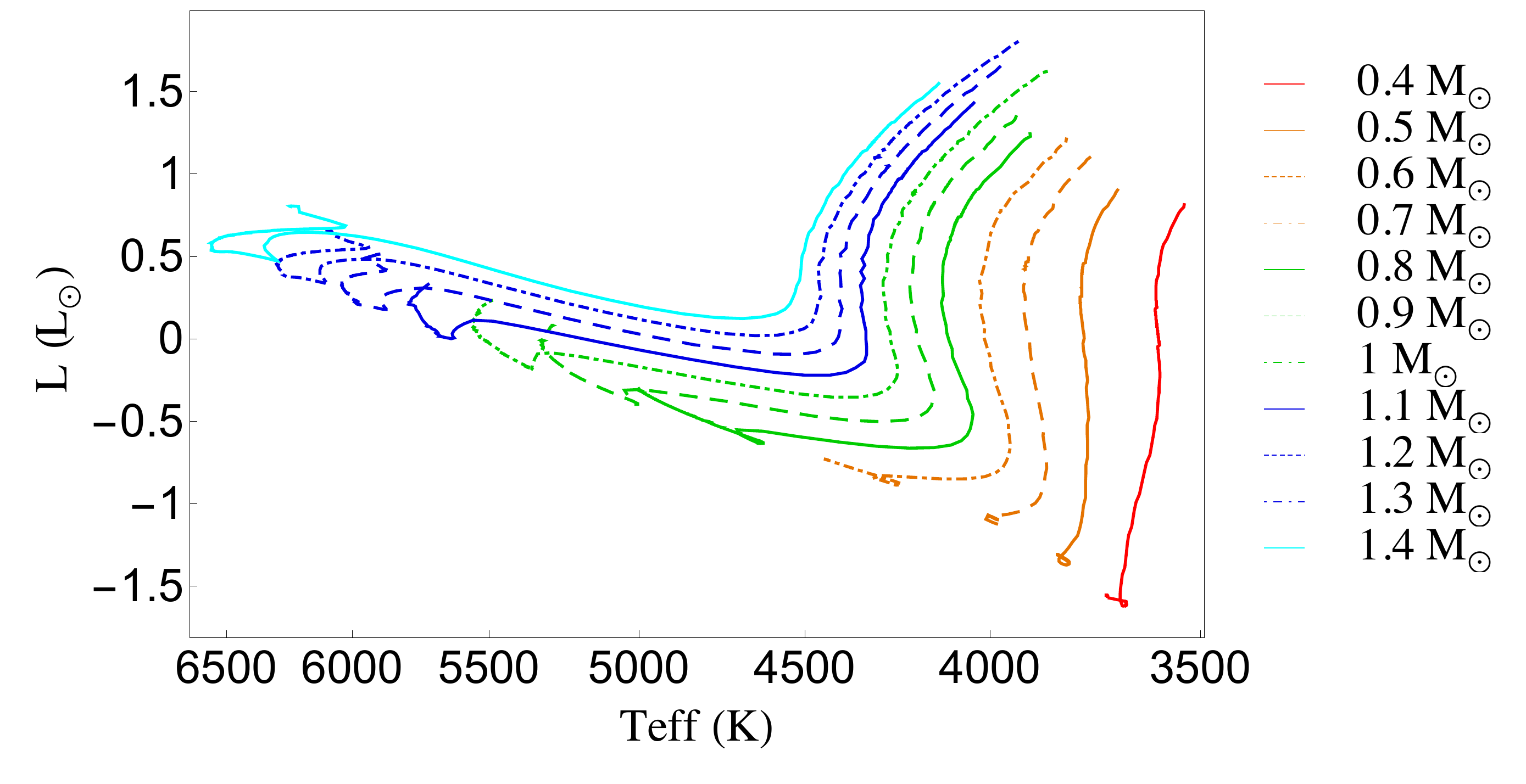}\quad\quad\quad
\includegraphics[width=0.35\linewidth]{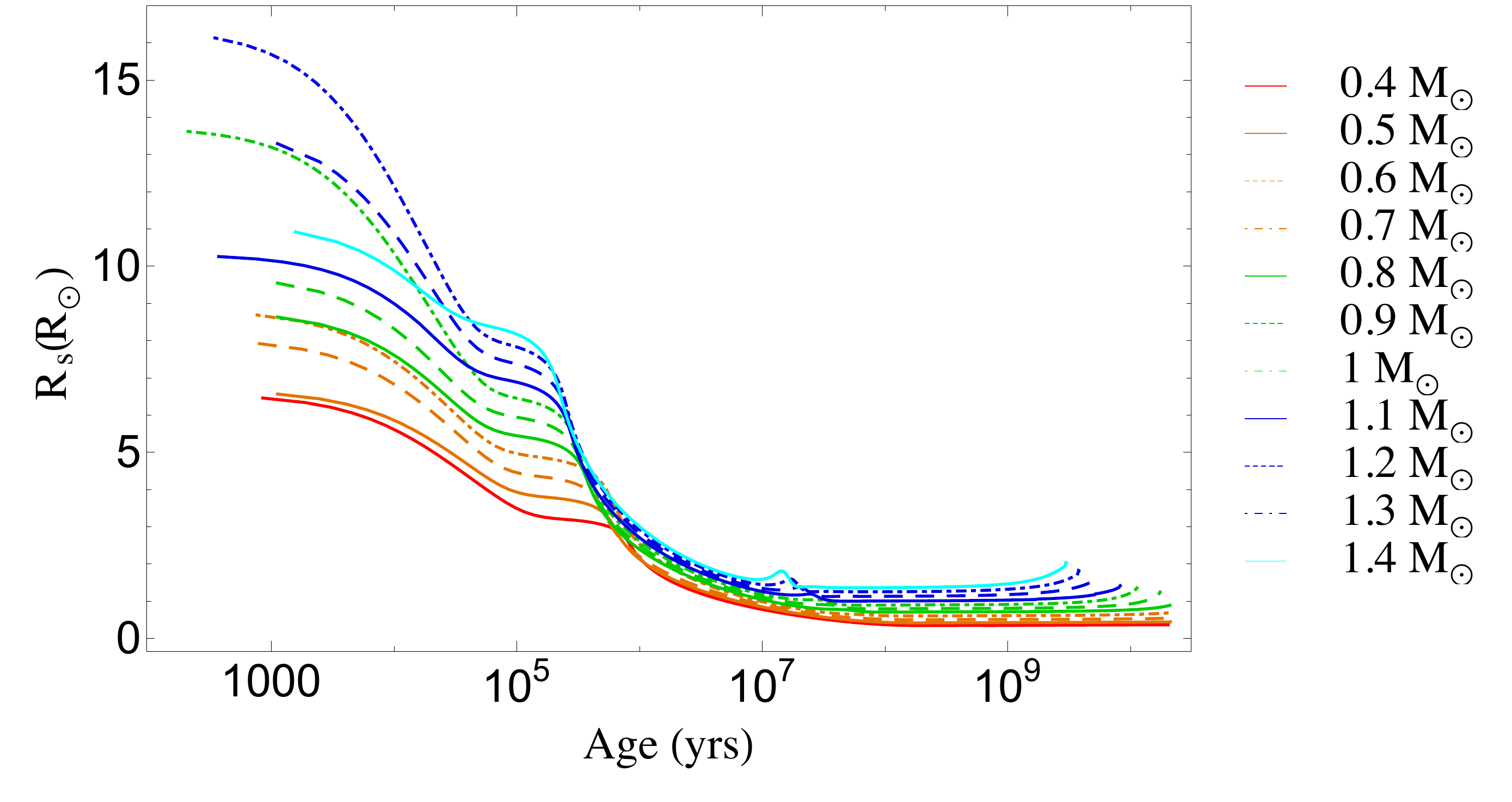}\\
\includegraphics[width=0.35\linewidth]{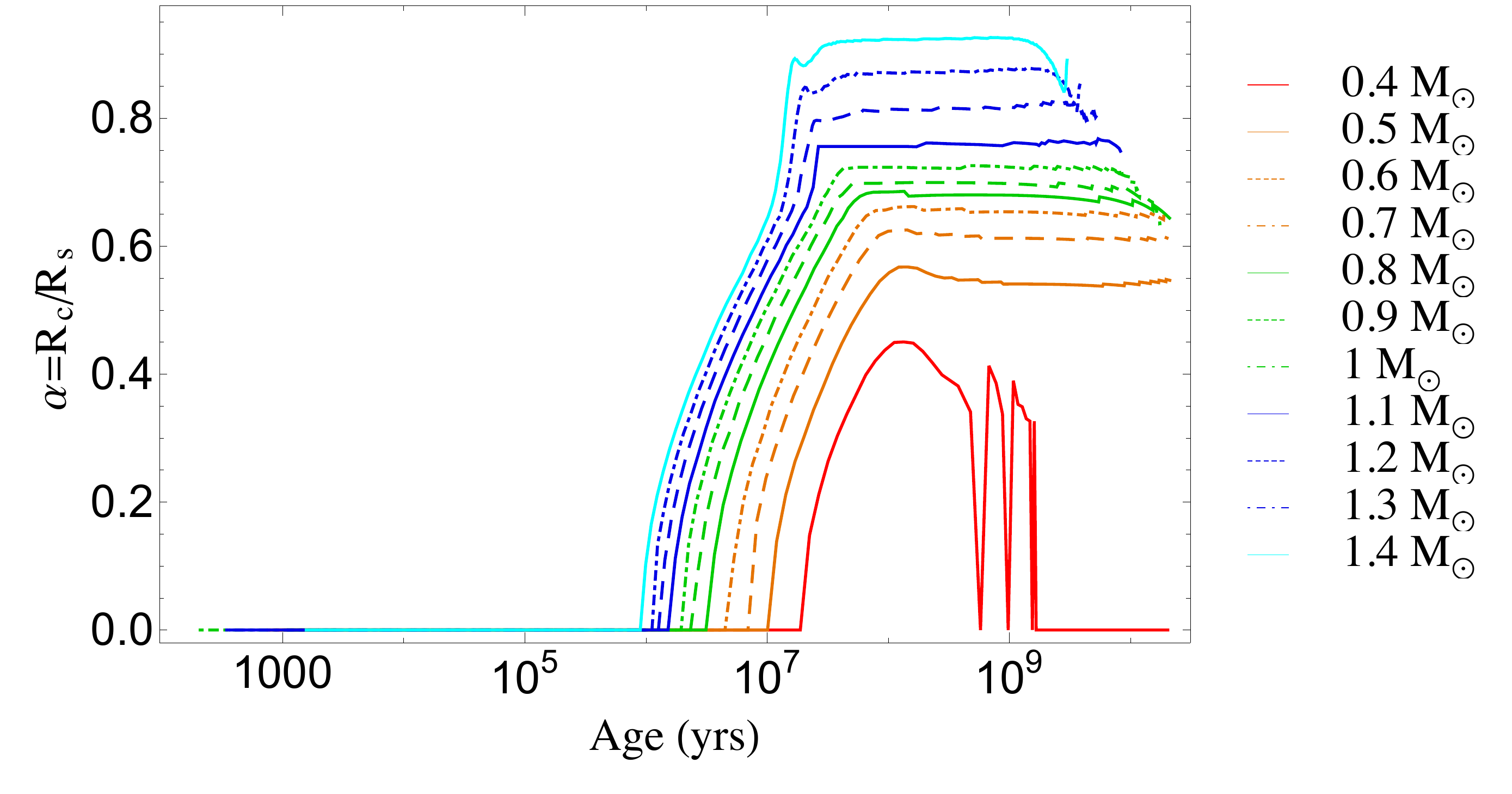}\quad\quad\quad
\includegraphics[width=0.35\linewidth]{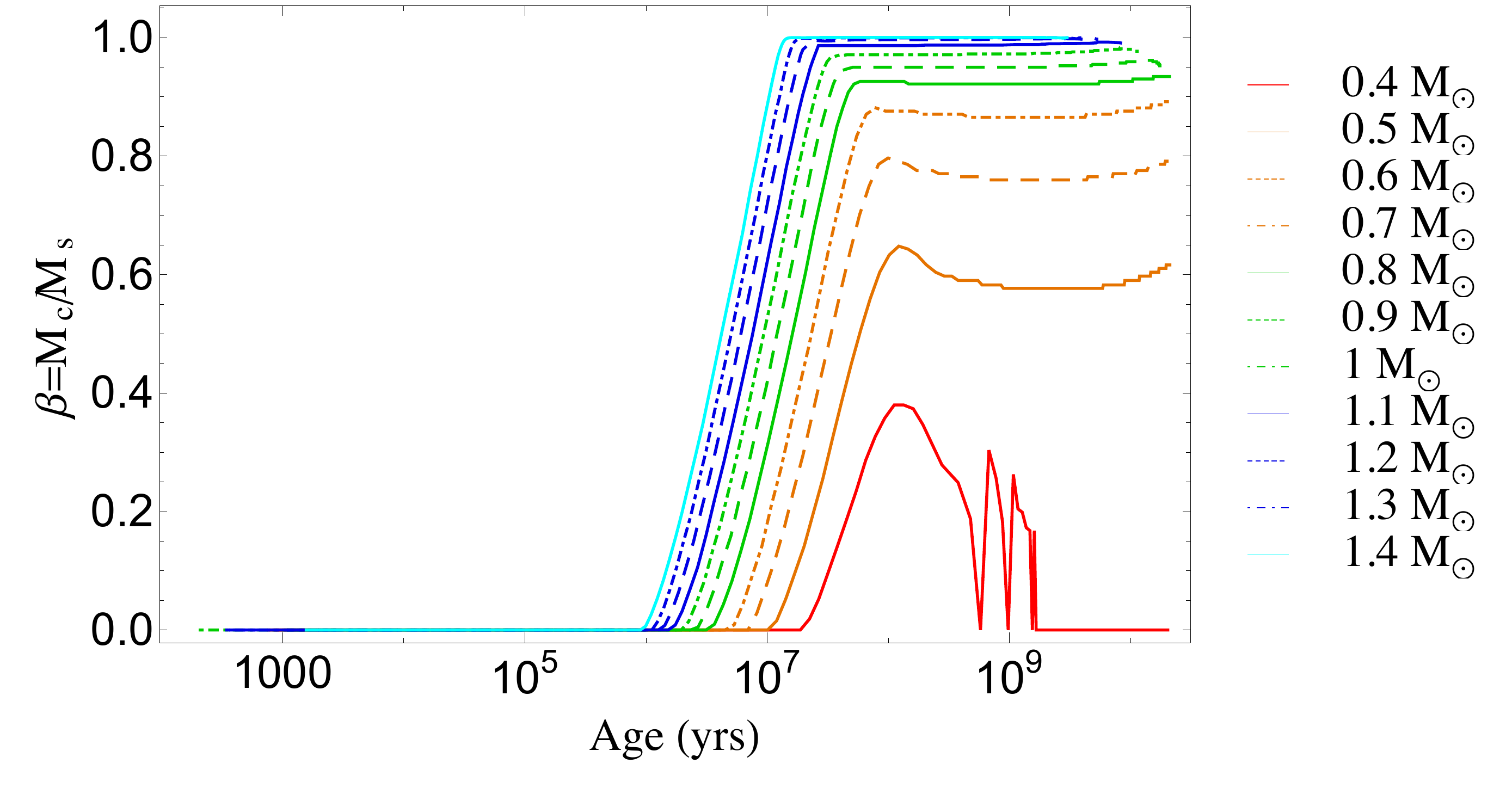}
\end{center} 
\caption{{\bf Top-left:} track of evolution of stars from $0.4$ to $1.4M_{\odot}$ in the Hertzsprung-Russell diagram that gives the luminosity ($L$) as a function of effective temperature ($T_{\rm eff}$) (red, orange, green, dark blue, cyan lines correspond to M, K, G, F and A-type stars respectively). {\bf Top-right:} Evolution of the stellar radius $R_{\rm s}$ of stars from $0.4$ to $1.4M_{\odot}$ as a function of time. {\bf Bottom-left:} Evolution of the radius aspect ratio $\alpha=R_{\rm c}/R_{\rm s}$ of stars from $0.4$ to $1.4M_{\odot}$ as a function of time. {\bf Bottom-right:} Evolution of the mass aspect ratio $\beta=M_{\rm c}/M_{\rm s}$ of stars from $0.4$ to $1.4M_{\odot}$ as a function of time.}
\label{Evol}
\end{figure*}

To analytically evaluate the frequency-averaged tidal dissipation in the convective envelope of low-mass stars, we here choose to adopt a simplified two-layer model as in \cite{Ogilvie2013} \citep[see also][and fig. \ref{SetUp}]{PZJ2014}. In this modeling, both the radiative core and the convective envelope are assumed to be homogeneous with respective constant densities $\rho_{\rm c}$ and $\rho_{\rm e}$ for the sake of simplicity. This model features a star A of mass $M_{\rm s}$ and mean radius $R_{\rm s}$ hosting a point-mass tidal perturber B of mass $m$ orbiting with a mean motion $n$. The convective envelope of A is assumed to be in moderate solid-body rotation with an angular velocity $\Omega$ so that $\epsilon^2 \equiv \left(\Omega/ \sqrt{\mathcal{G} M_{\rm s} / R_{\rm s}^3}\right)^2=\left(\Omega/\Omega_{\rm c}\right)^2 \ll 1$\footnote{In this regime, the centrifugal acceleration, which scales as $\Omega^{2}$ is neglected.}, where $\Omega_{\rm c}$ {is the critical angular velocity} and ${\mathcal G}$ is the gravitational constant. It surrounds the radiative core of radius $R_{\rm c}$ and mass $M_{\rm c}$. 

Tidal dissipation in the external convection zone of A originates from the excitation by B of inertial waves, which have the Coriolis acceleration as restoring force. They are damped by the turbulent friction, which is modeled using a turbulent viscosity \citep[see][and references therein]{OL2012}. Its analytical evaluation in our two-layer model was conducted by \cite{Ogilvie2013} who assumed an homogeneous and incompressible convective envelope\footnote{This assumption corresponds to inertial waves with shorter wavelength than the characteristic length of the variation of density.} surrounding an homogeneous stable fluid core where no dissipation occurs. The solutions of the system of dynamical equations for the envelope written in the co-rotating frame are separated into a non-wavelike part (with subscripts $_{\rm nw}$), which corresponds to the immediate hydrostatic adjustment to the external tidal potential ($U$), and a wavelike part (with subscript $_{\rm w}$) driven by the action of the Coriolis acceleration on the non-wavelike part {(see Appendix \ref{Appendix1} and \cite{Ogilvie2013})}. The kinetic energy of the wavelike part of the solution can be derived without solving the whole system of equations, thanks to an impulsive calculation. It is dissipated after a finite time and allows us to compute the frequency-averaged tidal dissipation {given in \cite{Ogilvie2013} (eq. B3)}\footnote{In the calculation, the weighted frequency integral reduces to $\omega \in \left[-2 \Omega,2\Omega\right]$ because higher-frequency acoustic waves are filtered out.}:
\begin{eqnarray}
\lefteqn{\left<{\mathcal D}\right>_{\omega}=\int^{+\infty}_{-\infty} \! {\rm Im} \left[k_2^2(\omega)\right] \,\frac{\mathrm{d}\omega}{\omega} = \frac{100 \pi}{63} \epsilon^2 \left(\frac{\alpha^5}{1-\alpha^5}\right)\left(1-\gamma\right)^2}\\
&&\times\left(1-\alpha\right)^4\left(1+2\alpha+3\alpha^2+\frac{3}{2}\alpha^3\right)^2\left[1+\left(\frac{1-\gamma}{\gamma}\right)\alpha^3\right]\nonumber\\
&&\times\left[1+\frac{3}{2}\gamma+\frac{5}{2\gamma}\left(1+\frac{1}{2}\gamma-\frac{3}{2}\gamma^2\right)\alpha^3-\frac{9}{4}\left(1-\delta\right)\alpha^5\right]^{-2}\nonumber
\label{eq:imk22fogilvie}
\end{eqnarray}
with
\begin{equation}
\alpha=\frac{R_{\rm c}}{R_{\rm s}}\hbox{,}\quad\beta=\frac{M_{\rm c}}{M_{\rm s}}\quad\hbox{and}\quad\gamma=\frac{\rho_{\rm e}}{\rho_{\rm c}}=\frac{\alpha^3\left(1-\beta\right)}{\beta\left(1-\alpha^3\right)}<1.
\end{equation}
We introduce the tidal frequency $\omega=sn-m\Omega$ (with $s\in{\pmb Z}$) and the Love number $k_{l}^{m}$, associated to the $\left(l,m\right)$ component of the time-dependent tidal potential $U$ that corresponds to the spherical harmonic $Y_l^m$. It quantifies at the surface of the star A ($r=R_{\rm s}$) the ratio of the tidal perturbation of its self-gravity potential over the tidal potential. In the case of a dissipative fluid as in convective envelopes, it is a complex quantity that depends on the tidal frequency with a real part that accounts for the energy stored in the tidal perturbation while the imaginary part accounts for the energy losses \citep[e.g.][]{RMZ2012}. Note that ${\rm Im}\left[ k_l^m(\omega) \right]$ is proportional to ${\rm sgn}(\omega)$ and can be expressed in terms of the tidal quality factor $Q_l^m(\omega)$ or equivalently the tidal angle $\delta_l^m(\omega)$, which both depend on the tidal frequency: 
\begin{equation}
{Q_l^m(\omega)}^{-1} = {\rm sgn} (\omega)\,{\left| k_l^m(\omega) \right|}^{-1}\,{{\rm Im}\left[ k_l^m(\omega) \right]} = \sin\left[2 \, \delta_l^m(\omega)\right].
\end{equation}
We here choose to consider the simplest case of a coplanar system for which the tidal potential ($U$) reduces to the component $(l=2,m=2)$  as well as the quadrupolar response of A. To unravel the impact of the variation of stellar internal structure, we introduce a {first} frequency-averaged dissipation at fixed {normalized} angular velocity
\begin{equation}
\left<{\mathcal D}\right>_{\omega}^{\Omega}=\epsilon^{-2}\left<{\mathcal D}\right>_{\omega}=\epsilon^{-2}\left<{\rm Im} \left[k_2^2(\omega)\right]\right>_{\omega},
\label{ADFOm}
\end{equation}
{which depends on $\alpha$ and $\beta$ only, and a second one
\begin{equation}
\left<{\hat{\mathcal D}}\right>_{\omega}^{\Omega}={\hat\epsilon}^{-2}\left<{\mathcal D}\right>_{\omega}\!=\!\left(\frac{M_{\rm s}}{M_{\odot}}\right)^{-1}\left(\frac{R_{\rm s}}{R_{\odot}}\right)^{3}\left<{\mathcal D}\right>_{\omega}^{\Omega},
\label{MoyEvol}
\end{equation}
where ${\hat \epsilon}^{2}\equiv \left(\Omega/ \sqrt{\mathcal{G} M_{\odot} / R_{\odot}^3}\right)^2=\left(\Omega/\Omega_{\rm c}^{\odot}\right)^2$ ($M_{\odot}$, $R_{\odot}$ and $\Omega_{\rm c}^{\odot}$ being the solar mass, radius and critical angular velocity), that allows us to take also into account the dependences on the radius ($R_{\rm s}$) variations and on the mass ($M_{\rm s}$). Note that these quantities being averaged in frequency, the complicated frequency-dependence of the dissipation in a spherical shell \citep[][]{OL2007} is filtered out. The dissipation at a given frequency could thus be larger or smaller than its averaged value ($\left<{\mathcal D}\right>_{\omega}$) by several orders of magnitude. Moreover, $\left<{\mathcal D}\right>_{\omega}$ constitutes a lower bound of the typical dissipation in the whole star since it is also necessary to take into account the damping of tidal gravito-inertial waves in radiation zones \citep[][]{Ivanovetal2013}.}

\section{Tidal dissipation along stellar evolution}

\subsection{Grids of models for low-mass stars}

To study the variation of tidal dissipation in the convective envelope of low-mass stars along their evolution in the theoretical framework presented above we need to compute the {stellar radius ($R_{\rm s}$) and the} mass and aspect ratios (respectively $\alpha$ and $\beta$) as functions of time. We use grids of stellar models from $0.4$ to $1.4M_{\odot}$ for a metallicity $Z=0.02$ computed by \cite{Siessetal2000} using the STAREVOL code. These stellar models solve, at the zeroth-order, the hydrostatic and energetic spherical balances using the relevant equation of state and description of stellar atmospheres \citep[a complete description of the STAREVOL code is given in][]{Siessetal2000}. They allow us to follow the evolution of stars and of their structure along their evolutionary tracks in the Hetzsprung-Russell diagram (Fig. \ref{Evol}, top-left panel). In this work, we choose to focus on the Pre-Main-Sequence, Main-Sequence and Sub-Giant phases of evolution (hereafter reported as PMS, MS and SG). In Fig. \ref{Evol} (top-right panel), the variation of the radius of the star ($R_{\rm s}$) is plotted as a function of time. For every stars, we identify the three well-known phases: i) during the PMS, stars are in contraction and their radius ($R_{\rm s}$) decreases; ii) on the MS it reaches an almost constant value until the beginning of the SG phase; iii) on the SG phase, stars' envelope begin to expand and $R_{\rm s}$ increases. As this has been seen in Sec. 2., the important structural properties to compute for tidal dissipation are {also} the mass and aspect ratios relative to the radiative core on which tidal inertial waves propagating in the convective envelope can reflect, leading potentially to wave attractors where strong dissipation may occur \citep[e.g.][]{OL2007,GL2009}. In Fig. \ref{Evol}, we thus plot the temporal evolution of $\alpha$ and $\beta$ for the different stellar masses studied here (respectively in bottom-left and bottom-right panels). After a first phase where the star is completely convective on the Hayashi phase, the central radiative core is growing both in radius and mass along the PMS. Next, for stars with $M_{\rm s}\ge0.5M_{\odot}$, both $\alpha$ and $\beta$ reach a value on the Zero-Age-Main-Sequence (ZAMS) that stays almost constant along the MS. For example, in the case of a $1M_{\odot}$ star we recover the usual solar values $\alpha\approx0.71$ and $\beta\approx 0.98$. Both $\alpha$ and $\beta$ increase with $M_{\rm s}$ on the MS. Finally, when stars reach their SG phase, both $\alpha$ and $\beta$ decrease because of the simultaneous extension of their convective envelope and contraction of their radiative core.

\subsection{Time evolution of tidal dissipation as a function of stellar mass and evolutionary stage}

\begin{figure}[!t]
\begin{center}
\includegraphics[width=0.7\linewidth]{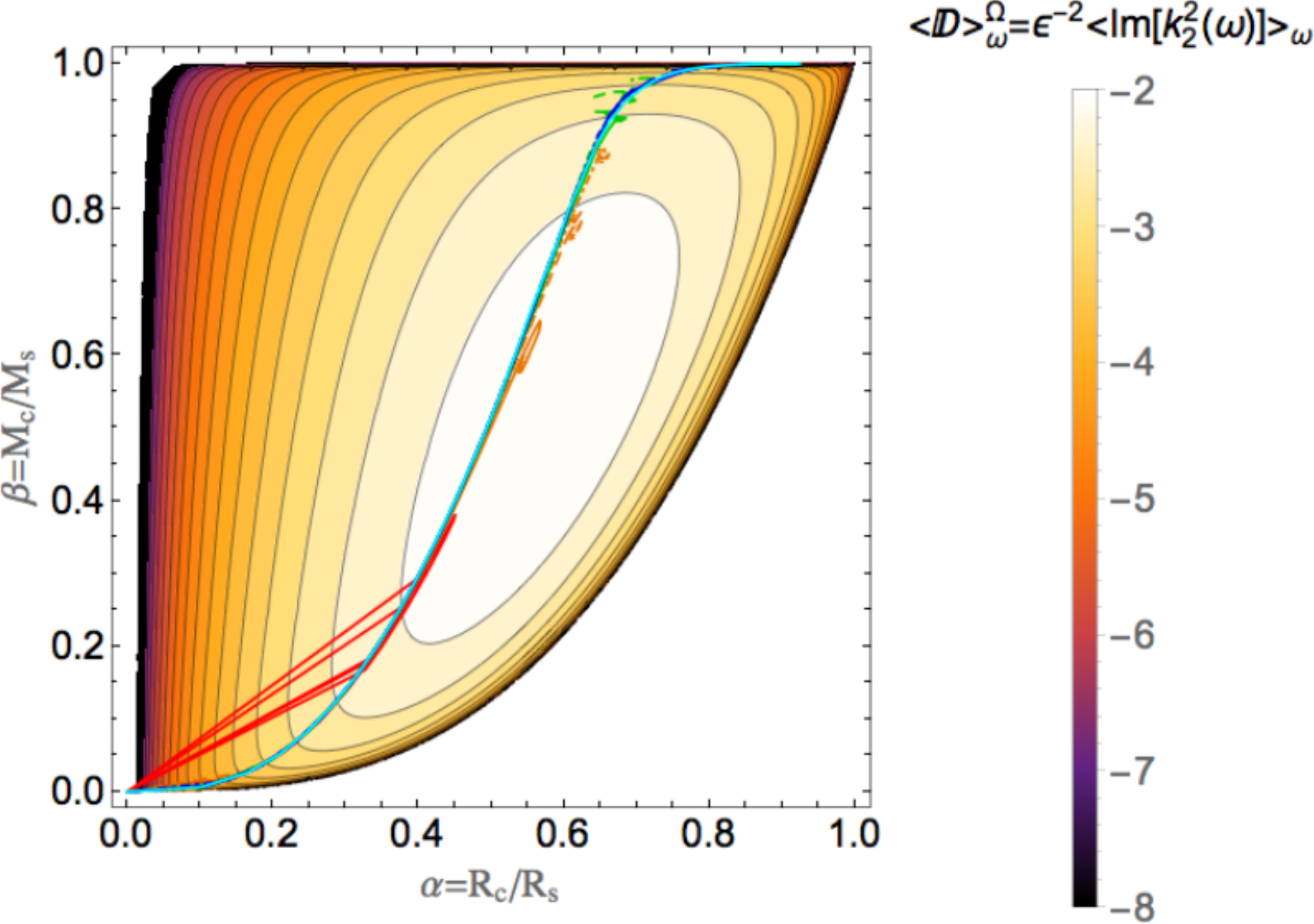}
\end{center} 
\caption{Variation of  $\left<{\mathcal D}\right>_{\omega}^{\Omega}={\epsilon}^{-2}\left<{\rm Im} \left[k_2^2(\omega)\right]\right>_{\omega}$ as a function of aspect and mass ratios ($\alpha$ and $\beta$ respectively) in color scales. Evolutionary tracks of stars from $0.4$ to $1.4M_{\odot}$ in the $\left(\alpha,\beta\right)$ plane (the colors are the same as in Fig \ref{Evol}).}
\label{TidalTrack}
\end{figure}

\begin{figure*}[!t]
\begin{center}
\includegraphics[width=0.4\linewidth]{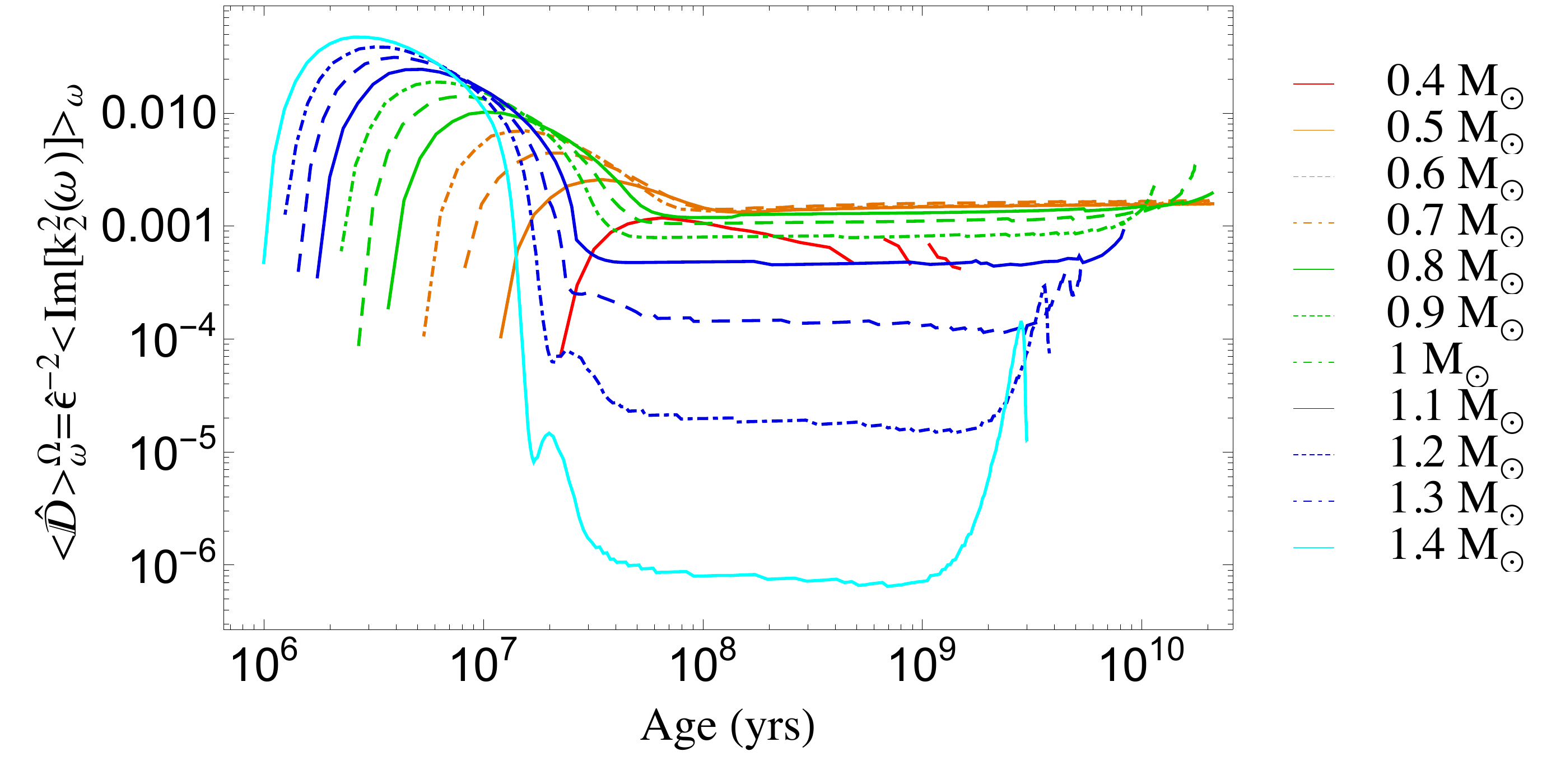}\quad\quad\quad
\includegraphics[width=0.4\linewidth]{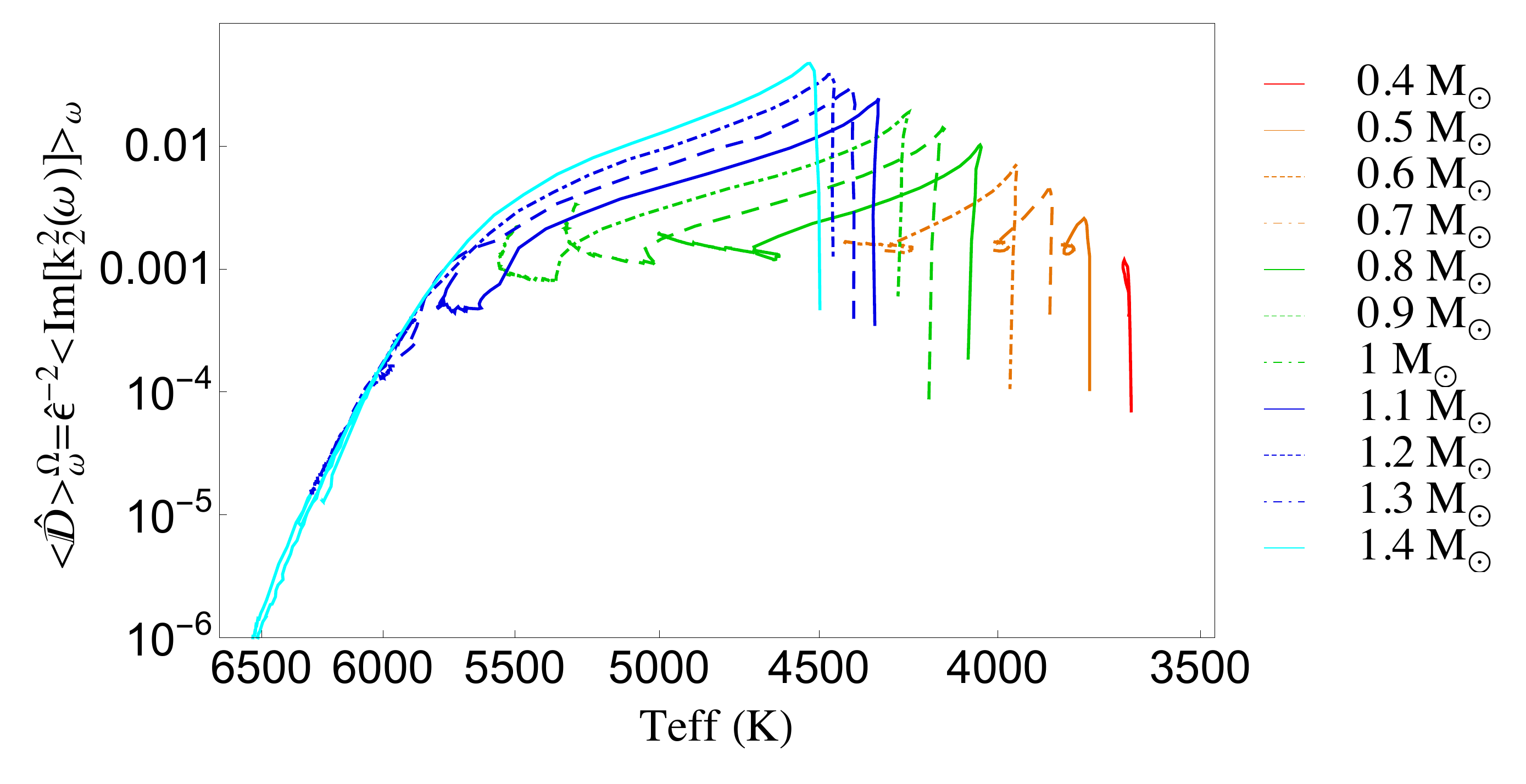}
\end{center} 
\caption{Evolution of the frequency-averaged tidal dissipation at fixed normalized angular velocity, 
$\left<{\hat{\mathcal D}}\right>_{\omega}^{\Omega}={\hat\epsilon}^{-2}\left<{\rm Im} \left[k_2^2(\omega)\right]\right>_{\omega}$, as a function of time (left panel) and effective temperature (right panel) for stellar masses ($M_{\rm s}$) from $0.4$ to $1.4M_{\odot}$ (the colors are the same as in Fig \ref{Evol}).}
\label{TidalAmplitude}
\end{figure*}

The phases of the structural evolution of low-mass stars being recalled, we now have to compute the corresponding evolution of the dissipation of the kinetic energy of tidal inertial waves in their convective envelope as a function of time. In Fig. \ref{TidalTrack}, we thus represent: i) the intensity of the {first} frequency-averaged dissipation at fixed {normalized} angular velocity ($\left<{\mathcal D}\right>_{\omega}^{\Omega}$; see Eq. \ref{ADFOm}) as a function of $\alpha$ and $\beta$ {only} in color scales and ii) the position of the couples $\left(\alpha,\beta\right)$ along the evolution of stars for stellar masses from $0.4$ to $1.4M_{\odot}$. On one hand, $\left<{\mathcal D}\right>_{\omega}^{\Omega}$ presents a maximum in an island region around $(\alpha_{\rm max}\approx0.571,\beta_{\rm max}\approx0.501)$. On the other hand, the stars of different masses have a different evolution of their internal tidal dissipation in their envelope because of their different structural evolution. First, all stars have an increase of $\left<{\mathcal D}\right>_{\omega}^{\Omega}$ during the beginning of their PMS until they reach a position in the diagram close to $\left(\alpha_{\rm max},\beta_{\rm max}\right)$ because of the formation and the growth of their radiative core. Then, $\left<{\mathcal D}\right>_{\omega}^{\Omega}$ has an evolution that is a direct function of stellar mass. For $0.4M_{\odot}$ M-type stars, the radiative core has an evolution where it finally disappears. Therefore, the configuration converges on the MS to the case of a fully convective star with the weak dissipation of normal inertial modes \citep{Wu2005}. For stars from $0.5M_{\odot}$ to $1.4M_{\odot}$ masses, the evolution is different. After they have reached their position closest to $\left(\alpha_{\rm max},\beta_{\rm max}\right)$ in the diagram, they evolve to a position $\left(\alpha_{\rm MS},\beta_{\rm MS}\right)$ where they stay during almost the whole MS because of the weak evolution of these quantities. Since $\alpha_{\rm MS}$ and $\beta_{\rm MS}$ are increasing functions of the stellar mass, $\left<{\mathcal D}\right>_{\omega}^{\Omega}$, which is almost constant on the MS, decreases from K to A-type stars. {However, the evolution of $R_{\rm s}$ must also been taken into account. Therefore,} the evolution of $\left<{\hat{\mathcal D}}\right>_{\omega}^{\Omega}$ (see Eq. \ref{MoyEvol}) as a function of time (left-panel) and of the effective temperature (right-panel) are given in Fig. \ref{TidalAmplitude} {taking into account the simultaneous variations of $\alpha$, $\beta$, and $R_{\rm s}$}. We recover the two phases of the evolution of the dissipation. On the PMS, it increases towards the maximum value{, which grows with stellar mass}, corresponding to the region around $\left(\alpha_{\rm max},\beta_{\rm max}\right)$ for all stellar masses. The time coordinate of this maximum is smaller if stellar mass is higher because of the corresponding shorter life-time of the star. Then, the dissipation decreases to reach rapidly its almost constant value on the MS. As already pointed out before and expected from observational constraints \citep[e.g.][]{Albrechtetal2012} and previous theoretical works \citep{OL2007,BO2009}\footnote{In the cases of F and A-type stars, a convective core is present in addition to the envelope. We neglect its action assuming it corresponds to the regime of weak dissipation in a full sphere.}, it decreases with stellar mass from {\bf $0.6M_{\odot}$} K-type to A-type stars by several orders of magnitude ({$\approx3$ between $0.6$} and $1.4M_{\odot}$) because of the variation of the thickness of the convective envelope, which becomes thinner. Finally, for higher-mass stars, we can see a final increase of $\left<{\hat{\mathcal D}}\right>_{\omega}^{\Omega}$ because of the simultaneous extension of their convective envelope and contraction of their radiative core during their SG phase leading again the star towards the region of maximum dissipation in the $(\alpha,\beta)$ plane.\\ 

\subsection{The impact of the rotational evolution of stars}

It is also crucial to discuss consequences of the rotational evolution of low-mass stars. As we know from observational works and related modeling \citep[e.g.][and references therein]{GB2013,GB2015}, their rotation follows three main phases of evolution: i) first, stars are trapped in co-rotation with the surrounding circumstellar disk; ii) next, because of the contraction of stars on the PMS (e.g. Fig. \ref{TidalTrack}, top-left panel) their rotation (and thus $\epsilon$) increases; iii) finally, stars are braked on the MS because of the torque applied by pressure-driven stellar winds \citep[e.g.][and references therein]{Revilleetal2015} and $\epsilon$ decreases. As a detailed computation of stellar rotating models is out of the scope of the present work because of the complex internal and external mechanisms that must be taken into account, we will not here evaluate the variation of tidal dissipation because of the simultaneous evolution of stellar internal structure and rotation. However, from results obtained by \cite{GB2013,GB2015}, we can easily infer that $\left<{\mathcal D}\right>_{\omega}$ is increased during the PMS because both of the growths of the radiative core and of the angular velocity. From Fig. 5 of \cite{GB2015}, we can see that in the case of the $0.5M_{\odot}$ star these growths will be simultaneous. On the MS, $\left<{\mathcal D}\right>_{\omega}$ decreases because of the evolution of the structure of the star and of its braking by stellar winds. As demonstrated before, the value of $M_{\rm s}$ is determinant.

\section{Conclusions}

All low-mass stars have an increase of the frequency-averaged tidal dissipation for a fixed angular velocity in their convective envelope {for tidal frequencies lying within the range $\left[-2\Omega,2\Omega\right]$ so that inertial waves can be excited} until they reach a critical aspect and mass ratios close to ($\alpha_{\rm max},\beta_{\rm max}$) during the PMS. Next, it evolves on the MS to an asymptotic value that {becomes} maximum for $0.6M_{\odot}$ K-type stars and that decreases by several orders of magnitude with increasing stellar mass. Finally, {the rotational evolution of low-mass stars strengthens the importance of tidal dissipation during the PMS for star-planet and multiple star systems as pointed out before by \cite{Zahn1989}}.

In a near future, it would be important to improve the physics of dissipation in stellar convection zones by taking into account density stratification, differential rotation, magnetic field {and nonlinear effects and related possible instabilities} \citep[e.g.][]{BR2013,Schmitt2010,Favieretal2014}. Moreover, it will be necessary to {include results obtained for the dissipation of} tidal gravito-inertial waves in stellar radiative cores \citep{Ivanovetal2013} to get a complete picture \citep{Guillotetal2014}. Finally, it will be interesting to explore advanced phases of stellar evolution.

\begin{acknowledgements}
{The author is grateful to the referee, A. Barker, for his detailed review, which has allowed to improve the paper.} This work was supported by the Programme National de Plan\'etologie (CNRS/INSU) and the CoRoT-CNES grant at Service d'Astrophysique (CEA-Saclay). S.M. dedicates this article to J. Abrassart.
\end{acknowledgements}

\bibliographystyle{aa}  
\bibliography{Mathis2015} 

\begin{appendix}

\section{Dynamical equations for tidal flows in convective envelopes}\label{Appendix1}

The solutions of the system of dynamical equations for the envelope written in the co-rotating frame are separated into a non-wavelike part (with subscripts $_{\rm nw}$), which corresponds to the immediate hydrostatic adjustment to the external tidal potential ($U$), and a wavelike part (with subscript $_{\rm w}$) driven by the action of the Coriolis acceleration on the non-wavelike part:
\begin{equation}
\begin{cases}
\ddot{\mathbf{s}}_{\mathrm{nw}} = -\nabla W_{\mathrm{nw}} ,\\
h'_{\mathrm{nw}} + \Phi'_{\mathrm{nw}} + U = 0 ,\\
\rho_{\mathrm{nw}}' = -\nabla \cdot (\rho_{\rm e} \, \mathbf{s}_{\mathrm{nw}}), \\
\nabla^2 \Phi_{\mathrm{nw}}' = 4 \, \pi \, \mathcal{G} \, \rho_{\mathrm{nw}}',
\end{cases}
\hbox{and}\quad
\begin{cases}
\ddot{\mathbf{s}}_{\mathrm{w}} + 2 \, \Omega \, \mathbf{e}_z \times \dot{\mathbf{s}}_{\mathrm{w}} = -\nabla W_{\mathrm{w}} + \mathbf{f} ,\\
h'_{\mathrm{w}} = \Phi'_{\mathrm{w}} = \rho_{\mathrm{w}}' = 0, \\
\nabla \cdot (\rho_{\rm e} \, \mathbf{s}_{\mathrm{w}}) = 0, 
\end{cases}
\end{equation}
where $\mathbf{s}$ is the displacement, $\mathbf{e}_z$ the unit vector along the rotation axis, $h$ the specific enthalpy, and $\Phi$ is the self-gravitational potential of A. Primed variables denote an Eulerian perturbation in relation to the unperturbed state with unprimed variables. {Note that $U$ being of the first order in tidal amplitude, it appears in perturbation equations}. Finally, $W \equiv W_{\mathrm{nw}}+ W_\mathrm{w} = h' + \Phi' + U$ while $\mathbf{f} = -2 \, \Omega \, \mathbf{e}_z \times \dot{\mathbf{s}}_{\mathrm{nw}}$ is the acceleration driving the wavelike part of the solution.

\end{appendix}

\end{document}